\documentclass[11pt]{article}

\def\fun#1#2{\lower3.6pt\vbox{\baselineskip0pt\lineskip.9pt
\ialign{$\mathsurround=0pt#1\hfil##\hfil$\crcr#2\crcr\sim\crcr}}}

\newcommand{\bc}{\begin{center}}
\newcommand{\ec}{\end{center}}
\newcommand{\bd}{\begin{displaymath}}
\newcommand{\ed}{\end{displaymath}}
\newcommand{\be}{\begin{equation}}
\newcommand{\ee}{\end{equation}}
\newcommand{\ba}{\begin{array}}
\newcommand{\ea}{\end{array}}
\newcommand{\bt}{\begin{tabular}}
\newcommand{\et}{\end{tabular}}
\newcommand{\un}{\underline}

\newcommand{\lb}{\label}
\newcommand{\rf}{\ref}
\newcommand{\bp}{\begin{picture}}
\newcommand{\ep}{\end{picture}}

\begin{document}

\bc

{\LARGE\bf Anti-Grand Unification and Critical\\[2mm] Coupling
Universality}\\[5mm]
{\large\bf L.V.Laperashvili\footnote[1]{E-mail:
larisa@vxitep.itep.ru}}\\[5mm]
{\bf Institute of Theoretical and Experimental Physics,
B.Cheremushkinskay 25, Moscow 117259, Russia}\\[5mm]
{\large\bf H.B.Nielsen\footnote[2]{E-mail:
hbech@nbivms.nbi.dk}}\\[5mm]
{\bf The Niels Bohr Institute, Blegdamsvej 17, DK-2100, Copenhagen,
Denmark}

\ec
\begin{abstract}
The present work considers the phase transition between the confinement
and "Coulomb" phases in $U(1)$, $SU(2)$ and $SU(3)$-sectors of
Anti-grand unified theory described by regularized Wilson loop action.
It was shown the independence of the critical coupling constants of the
regularization method ("universality").
\end{abstract}

Standard model unifying QCD with Glashow--Salam--Weinberg electroweak
theory well describes all experimental results known today. Most
efforts to explain the Standard model are devoted to Grand
unification theory (GUT). The precision of the LEP--data allows to
extrapolate three coupling constants of the Standard model to high
energies with small errors and we are able to perform consistency
checks of the Grand unification theories.

In the Standard model based on the group
\be
SMG=SU(3)_c\otimes SU(2)_L\otimes U(1)  \lb{1}
\ee
the usual definitions of the coupling constants are used:
\be
\alpha_1=\frac{5}{3}\frac{\alpha}{\cos^2\theta_{\overline{MS}}},\quad
\alpha_2=\frac{\alpha}{\sin^2\theta_{\overline{MS}}},\quad
\alpha_3\equiv\alpha_S=\frac{g^2_S}{4\pi},    \lb{2}
\ee
where $\alpha$ and $\alpha_S$ are the electromagnetic and strong fine
structure constants, respectively. Using experimentally given parameters:

\be
\ba{ll}
\sin^2\theta_{\overline{MS}}(M_Z)=0.2316\pm0.0003,\\[2mm]
\alpha_S(M_Z)=0.118\pm0.003,\quad \quad \alpha^{-1}(M_Z)=127.9\pm0.1, \lb{3}
\ea
\ee
it is possible to extrapolate the experimental values of three
inverse running constants $\alpha^{-1}_i(\mu)$ to the Planck scale:
$\mu_{Pl}=1.22\cdot10^{19}$GeV.

The comparison of the evolutions of the inverses of the running
coupling constants to high energies in the Minimal Standard model
(MSM) (with one Higgs doublet) and in the Minimal Supersymmetric
Standard model (MSSM) (with two Higgs doublets) gives rise to the existence
of the grand unification point at
$\mu_{GUT}\sim10^{16}$GeV only in the case of MSSM (see Ref.\cite{1}).
This observation is true for a whole class of GUT's that break to the
Standard model group in one step, and which predict a "grand desert"
between the weak (low) and the grand unification (high) scales.
If grand desert indeed exists, and the supersymmetry is established
at future colliders then we shall eventually be
able to use the coupling constant unification to probe the new
physics near the unification and Planck scales.

Scenarios based on the Anti-grand unification theory ($AGUT$) was
developed in Refs.\cite{2}-\cite{5} as an alternative to GUT's.

Anti-grand unified theory suggests that at the Planck scale
$\mu_{Pl}$, considered as a fundamental scale, there exists the more
fundamental gauge group $G$, containing
$N_{gen}$ copies of the Standard model group $SMG$:
\be
G=SMG_1\otimes SMG_2\otimes \ldots\otimes
SMG_{N_{gen}}\equiv(SMG)^{N_{gen}},    \lb{4}
\ee
where the integer $N_{gen}$ designates the number of quark and lepton
generations .

The theory predicts and experiment confirms, that $N_{gen}=3$.
Subsequently, the fundamental gauge group G is:
\be
G=(SMG)^3=SMG_1\otimes SMG_2\otimes SMG_3 .      \lb{5}
\ee

The AGUT assumes that Nature seeks a special point -- the multiple
critical  point (MCP) where the group $G$ undergoes spontaneous
breakdown to the diagonal subgroup:
\be
G\to G_{diag.subgr.}=\left\{g, g, g\parallel g\in SMG\right\} \lb{6}
\ee
which is identified with the usual (lowenergy) group SMG.

This means that at the Planck scale the fine structure constants
$\alpha_Y\equiv\frac{3}{5}\alpha_1$, $\alpha_2$ and
$\alpha_3,$ as chosen by Nature, are just the ones
corresponding to the multiple critical point (MCP)
which is a point where all action parameter (coupling)
values meet in the phase diagram of the regularized Yang-Mills $(SMG)^3$
- gauge theory.

The extrapolation of the experimental values of the inverses
$\alpha^{-1}_{Y,2,3}(\mu)$ to the Planck scale $\mu_{Pl}$ by the
renormalization group formulas (under the assumption of a "desert" in
doing the extrapolation with one Higgs doublet) is shown in Fig.1
and gives us the following  result:
\be
\alpha^{-1}_Y(\mu_{Pl})=55.5;\quad\alpha^{-1}_2(\mu_{Pl})=49.5;\quad\alpha
^{-1}_3(\mu_{Pl})=54.        \lb{7}
\ee

\vspace{0cm}

Properties of Anti-grand unification can be studied by means of Monte
Carlo simulations, which indicate the existence of the multiple
critical point.

Using theoretical corrections to the Monte Carlo results on lattice,
Anti-grand unified theory gives the following predictions:\\
\bc
{\bf Table 1}\\
\ec

\bc
\begin{tabular}{|p{1.5cm}|p{4cm}|p{5cm}|}
\hline
Group & AGUT predictions & "Experiment" -- the extrapolation
                                of the SM results to the Planck
                                scale\\

\hline $U(1)$ & $\alpha^{-1}_Y(\mu_{MCP})=55\pm6$
&$\alpha^{-1}_Y(\mu_{Pl})\approx55.5$\\

\hline $SU(2)$ &
$\alpha^{-1}_2(\mu_{MCP})=49.5\pm3$
&$\alpha^{-1}_2(\mu_{Pl})\approx49.5$\\

\hline $SU(3)$&
$\alpha^{-1}_3(\mu_{MCP})=57\pm3$
&$\alpha^{-1}_3(\mu_{Pl})\approx54$\\

\hline
\end{tabular}
\ec
\vspace{0.5cm}
For $U(1)$ - gauge lattice theory the authors of Ref.\cite{6} have investigated
the behaviour of the effective fine structure constant in the vicinity of
the critical point and they have obtained:
\be
\alpha_{crit}\approx0.2.     \lb{9}
\ee

We gave put forward the calculations of the fine
structure constant in $U(1)$ - gauge theory, suggesting that the
modification of the action form might not change too much the
critical value of the effective coupling constant.

Instead of the lattice hypercubic regularization we have considered
rather new regularization using Wilson loop (nonlocal) action \cite{7}:
\be
S=\int
d^4x\int\limits^{\infty}_0d\log(\frac{R}{a})\beta(R)R^{-4}\sum\limits_{average}
Re Tr exp\left[i \oint_{C(R)} \hat A_{\mu}(x)dx^{\mu}\right]   \lb{10}
\ee
in approximation of circular Wilson loops $C(R)$ of radius $R\ge a$.
In the last equation we have the average $(\sum\limits_{average})$
over all positions and orientations of the Wilson loops $C(R)$ in
4-dimensional (Euclidean) space.

\begin{figure}[t]
\bc
\bp(100,55)
\put(0,-20){\includegraphics{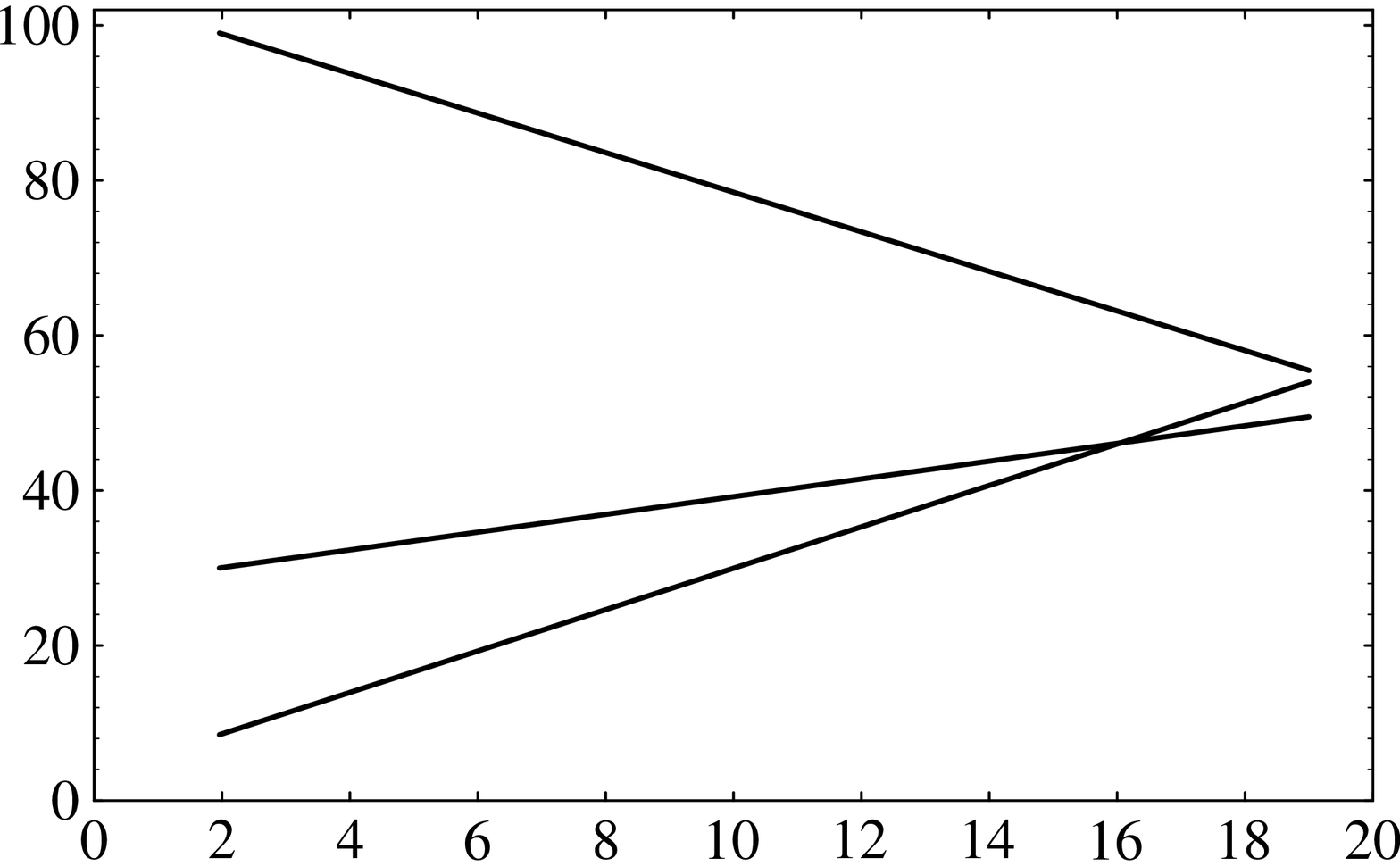}}
\put(-2,45){\llap{$\alpha^{-1}$}}
\put(21,40){$\alpha_Y^{-1}$}
\put(21,19){$\alpha_2^{-1}$}
\put(30,7){$\alpha_3^{-1}$}
\put(90,-3){$\log_{10}\mu$}
\ep
\ec
\caption{}
\end{figure}

With this action we have investigated the partition function, giving the free
energy $F$:
\be
Z=\int[D A_{\mu}]e^{-S}=e^{-F}.      \lb{12}
\ee
At the phase transition point the change of the free energy is equal to zero:
\be
\Delta F=0.  \lb{13}
\ee
This condition allows us to obtain the Balance equation between energy
and entropy:
\be
\Delta F=\Delta<S>-\Delta E_{entr}=0,  \lb{14}
\ee
where $<S>$ is the vacuum mean value of the action which plays role
of energy, and $E_{entr}$ is the entropy.

We have investigated the behaviour of $\Delta<S>$, which gave us the
following function:
\be
f_1(x)=\left(\frac{\sqrt{8}}{x}\right)^4,     \lb{15}
\ee
where $x=Rp_{cutoff}$ and $p_{cutoff}$ is the cutoff momentum value of the
gauge field $\hat A_{\mu}(p)$.

The second function was obtained from the change of the entropy
$(\Delta E_{entr})$:
\be
f_2(x)=4\pi \frac{1-J^2_1(x)-J^2_0(x)}{x^2} =
       4\pi h(x),  \lb{17}
\ee
and contains two Bessel functions $J_0(x)$ and $J_1(x)$.

The intersection of these two functions $f_1(x)$ and $f_2(x)$ at $x_{crit}$
corresponds to the Balance equation (\rf{14})
and gives rise to:\ $ x_{crit}\approx{2.57}$.

The critical value of the fine structure constant for the regularized
$U(1)$ - gauge theory is related with the value of $h(x_{crit})$ in
the following way:
\be
\alpha^{-1}_{crit}-\alpha^{-1}_{max}=\frac{1}{2h(x_{crit})}  \lb{19}
\ee
where $\alpha_{\max}\approx0.26$. According to Eq.(\rf{19})
the critical value of the fine structure constant is:
\be
\alpha_{crit}\approx{0.204}.     \lb{21}
\ee
The result (\rf{21}) confirms the Monte Carlo simulation result (\rf{9}) on
the  lattice.

The further investigations of $SU(2)\otimes U(1)$ and $SU(3)$ gauge
theories with the regularization using fermion string action confirm the
"universality" of the critical coupling constants.
Such an approximate regularization independence --
"\un{universality}" -- of the critical coupling constants is needed
for the fine structure constant predictions claimed  from Anti-grand
unified theory \cite{2}-\cite{5}.

Financial support of INTAS (grants INTAS-93-3316-ext and
INTAS-RFBR-95-0567) is gratefully acknowledged.

\end{document}